\documentclass{article}
\usepackage{import}
\usepackage{booktabs}
\usepackage{tablefootnote}
\usepackage{amsfonts}
\usepackage{amssymb}
\usepackage{caption} 
\usepackage{spconf,amsmath,graphicx}
\usepackage{subcaption}

\title{Two-Stage Augmentation and Adaptive CTC fusion for improved Robustness of Multi-Stream End-to-End ASR}
\name{Ruizhi Li$^1$, Gregory Sell$^{1,2}$, Hynek Hermansky$^{1,2}$}
\address{  $^1$Center for Language and Speech Processing, The Johns Hopkins University, USA\\
  $^2$Human Language Technology Center of Excellence, The Johns Hopkins University, USA}

\begin{document}
\ninept
\maketitle
\begin{abstract}


Performance degradation of an Automatic Speech Recognition (ASR) system is commonly observed when the test acoustic condition is different from training. 
Hence, it is essential to make ASR systems robust against various environmental distortions, such as background noises and reverberations.
In a multi-stream paradigm, improving robustness takes account of handling a variety of unseen single-stream conditions and inter-stream dynamics.
Previously, a practical two-stage training strategy was proposed within multi-stream end-to-end ASR, where Stage-2 formulates the multi-stream model with features from Stage-1 Universal Feature Extractor (UFE).
In this paper, as an extension, we introduce a two-stage augmentation scheme focusing on mismatch scenarios:
Stage-1 Augmentation aims to address single-stream input varieties with data augmentation techniques; 
Stage-2 Time Masking applies temporal masks on UFE features of randomly selected streams to simulate diverse stream combinations. 
During inference, we also present adaptive Connectionist Temporal Classification (CTC) fusion with the help of hierarchical attention mechanisms.
Experiments have been conducted on two datasets, DIRHA and AMI, as a multi-stream scenario. 
Compared with the previous training strategy, substantial improvements are reported with relative word error rate reductions of $29.7-59.3\%$ across several unseen stream combinations.

\end{abstract}

\section{Introduction}
\label{sec:intro}

The multi-stream paradigm of speech processing has been an active research area, in which parallel information sources are simultaneously considered for knowledge fusion.  
A robust fusion strategy is crucial to reliably address a variety of scenarios with different dynamics across streams. 
As one inspiration, the idea of parallel processing in human auditory systems has successfully motivated developments of various multi-stream frameworks in hybrid ASR \cite{hermansky2013multistream, hermansky2018coding, mallidi2018practical, mallidi2016novel}. For instance, multi-band acoustic modeling \cite{mallidi2018practical, mallidi2016novel} was proposed to improve noise robustness for a speech recognizer. 
Performance measures were introduced to select the most informative source in spatial acoustic scenes for hearing aids \cite{meyer2016performance}  or determine the quality of the model outputs \cite{li2019pm}.
Multi-modal approaches combined visual \cite{palaskar2018end} or symbolic \cite{renduchintala2018multi} inputs together with speech signals to improve speech recognition.
This work concentrates on the setting of multiple far-field microphone arrays, e.g., meeting rooms or domestic scenarios.  
The common methods of combing multiple arrays in conventional ASR are posterior combination \cite{wang2018stream, xiong2018channel}, ROVER \cite{fiscus1997post}, distributed beamformer \cite{yoshioka2019meeting}, and selection based on Signal-to-Noise/Interference Ratio (SNR/SIR) \cite{du2018theustc}. 

The multi-stream end-to-end framework was present in previous studies \cite{li2019multi, wang2019stream}, in which the MEM-Array model was introduced for multi-array applications.
It is a single neural network that takes multiple inputs and directly outputs word/letter sequences. 
This framework was proposed based on a joint CTC/Attention E2E scheme \cite{kim2016joint_icassp2017,hori2017advances,watanabe2017hybrid}, where each stream is characterized by a separate encoder and CTC network. 
A Hierarchical Attention Network (HAN) \cite{wang2019stream, li2018multiencoder} acts as a fusion component to dynamically guide the system towards streams carrying more discriminative information.
A practical two-stage training strategy was introduced later in \cite{li2020practical}.
In Stage-1, an Universal Feature Extractor (UFE) is optimized without requiring parallel data; Stage-2 formulates a multi-stream model directly on the UFE features with focus on solely training the HAN component. 

The previous two-stage training strategy \cite{li2020practical} offers a promising direction to further improve the robustness of multi-stream systems. It involves augmentation of training data, with an emphasis on single-stream variations in Stage-1 and inter-stream dynamics in Stage-2. 
Moreover, in \cite{li2020practical}, pre-defined equal CTC contributions during inference can potentially confuse the decoding procedure, especially when acoustic conditions among streams are dramatically different. 

In this paper, we present a two-stage augmentation scheme and adaptive CTC fusion targeting the aforementioned situations. 
The proposed techniques have the following highlights:
\begin{enumerate}
    \item Stage-1 Augmentation aims to train a well-generalized encoder so that the resulting UFE features could be robust against different unseen stream conditions.
    Both online augmentation (SpecAugment \cite{Park2019SpecAugmentAS}) and offline augmentation approaches are explored. Stage-2 Time Masking applies temporal masks on the UFE features. It provides a simple online augmentation technique to create inter-stream dynamics. 
    \item Adaptive CTC fusion applies the stream fusion vector to the CTC networks in the decoding step. CTC contributions then change dynamically depending on the HAN component, instead of the previous approach of pre-fixed weights.
\end{enumerate}





\section{Multi-Stream End-to-End Framework}
In this section, we review the MEM-Array model, one representative framework of the multi-stream approach with focus on far-field microphone arrays. 
An efficient two-stage training strategy is also discussed. 

\subsection{MEM-Array Model}

An end-to-end ASR model addressing general multi-stream scenarios was proposed in \cite{li2019multi} within the joint CTC/Attention architecture. 
As one realization of the multi-stream approach, the MEM-Array model can take as input parallel streams from several distant microphone arrays. 
We denote a $T^{(i)}$-length sequence of $D$-dimensional speech vectors as $X^{(i)}=\{\textbf{x}_{t}^{(i)}\in \mathbb{R}^{D}|t = 1,2,...,T^{(i)}\}$, where superscript $i\in \{1,...,N\}$ is the index for $i$-th stream.  
The MEM-Array model directly maps $N$ information sources, $X = \{X^{(1)}, X^{(2)}, ... ,X^{(N)}\}$, into an $L$-length label sequence, $C=\{c_{l}\in \mathcal{U}|l = 1,2,...,L\}$. 
Here $\mathcal{U}$ is a set of distinct labels. 

In the MEM-Array model, multiple microphone arrays are activated by separate encoders with identical architectures to capture diverse information. 
The $\textrm{Encoder}^{(i)}$ operates on the acoustic sequence $X^{(i)}$ to extract a set of higher-level feature representations $H^{(i)}=\{\textbf{h}^{(i)}_{1},..., \textbf{h}^{(i)}_{\lfloor T^{(i)}/s\rfloor}\}$, where
$s$ is the subsampling factor defined by the encoder architecture. 

Two levels of attention mechanisms are designated to combine the different views.
A frame-level attention mechanism is assigned to each encoder to obtain the stream-specific speech-label alignment. 
A location-based attention network \cite{NIPS2015_5847} is applied to compute the letter-wise context vector $\textbf{r}_{l}^{(i)}$ for stream $i$:
\begin{equation}
\textbf{r}_{l}^{(i)}={\sum}_{t=1}^{\lfloor T^{(i)}/s\rfloor}a_{lt}^{(i)}\textbf{h}_{t}^{(i)},
\end{equation}
where ${a}^{(i)}_{lt}$ is the attention weight, a soft-alignment of $\textbf{h}^{(i)}_t$ for output $c_{l}$.
A hierarchical stream-level attention mechanism then handles different dynamics across the streams. 
The fusion context vector $\textbf{r}_l$ is computed in a content-based attention network \cite{NIPS2015_5847}:
\begin{equation}
\textbf{r}_{l}={\sum}_{i=1}^{N}\beta_{l}^{(i)}\textbf{r}_{l}^{(i)},
\end{equation}
\begin{equation}
\beta_{l}^{(i)}=\textrm{HierarchicalAttention}(\textbf{q}_{l-1}, \textbf{r}_l^{(i)}), i\in\{1, ..., N\}.
\end{equation}
The Softmax output $\beta_{l}^{(i)}$ represents a stream-level attention weight for stream $i$ of letter prediction $c_{l}$.

Moreover, a separate CTC network is designated for each encoder. Per-encoder CTC modules have pre-defined equal contributions for joint training and decoding. 
In the beam search, the CTC prefix score \cite{watanabe2017hybrid, graves2008supervised} $\alpha_\textrm{ctc}(h)$ of hypothesized sequence $h$ is as follows: 
\begin{equation}
\label{eq:ctc}
    \alpha_\textrm{ctc}(h)=\frac{1}{N}{\sum}_{i=1}^{N}\alpha_{\textrm{ctc}^{(i)}}(h),
\end{equation}
where equal weight is assigned to each CTC network.

\begin{figure}[h!]
  \centering 
  \centerline{\includegraphics[width=9.5cm]{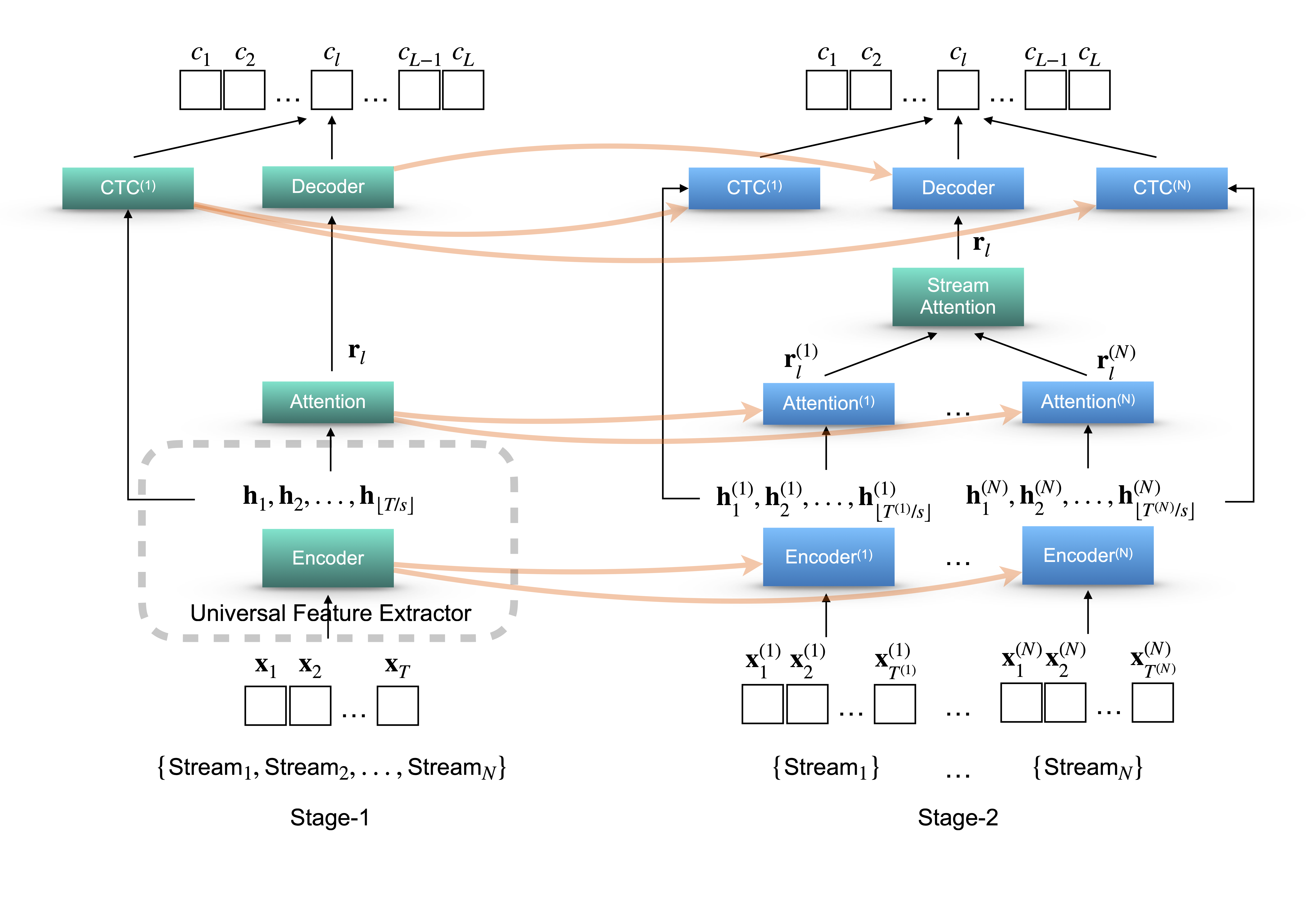}}
  \caption{Two-Stage Training Strategy \cite{li2020practical}. Color ``green'' indicates the components are trainable; Color ``blue'' means parameters of the components are frozen. }
  \label{fig:mema_2stage}
\end{figure}

\subsection{Two-Stage Training Strategy}

With an increasing number of streams (encoders) involved, jointly training a massive network requires substantial memory and vast amounts of parallel data. 
A two-stage training strategy was present in \cite{li2020practical} to tackle the aforementioned issues, depicted in Fig. \ref{fig:mema_2stage}.
This strategy resulted in performance improvements while efficiently scaling the training procedure.

Firstly, Stage-1 focuses on training a single-stream ASR model using various data with no presumption of parallel streams. 
The well-optimized encoder, which is referred to as Universal Feature Extractor (UFE), is used to further process acoustic frames from individual streams to generate UFE features, 
$\{H^{(1)}, H^{(2)}, ... , H^{(N)}\}$. 
In addition, byproducts in Stage-1, such as decoder, CTC and frame-level attention, are used for initialization in Stage-2. 
Secondly, Stage-2 formulates a multi-stream architecture directly operating on the UFE features as inputs with no highly-parameterized parallel encoders involved. 
We define the streams used for Stage-2 training as target streams.
With pre-trained components frozen during optimization, the model concentrates solely on the stream-level attention.

\section{Proposed Approaches for Robustness}
As an extension of the two-stage training strategy, in this section, we present a two-stage argumentation scheme and an adaptive CTC fusion to improve robustness of the MEM-Array model. 

\subsection{Two-Stage Augmentation}

Following the framework of the two-stage training strategy described in Fig. \ref{fig:mema_2stage}, the proposed two-stage augmentation scheme defines individual steps to simulate single-stream variations and inter-stream dynamics, respectively.

\subsubsection{Stage-1 Augmentation}

The goal of Stage-1 training is to obtain a set of UFE features with more discriminative power for Stage-2 prediction. 
With limited amount of data for target streams in Stage-2, data augmentation in Stage-1 is a strategy to create more data with a diverse set of conditions and also to involve audio from non-target arrays. 
In this work, we explore two approaches to improve training with data augmentation in the multi-array scenario:
\begin{itemize}
    \item SpecAugment \cite{Park2019SpecAugmentAS} is an online augmentation technique that degrades input on the fly in the training mini-matches. 
    It views the spectrogram as a visual representation, and modifies the spectrogram by warping it in the time direction and applying masks in frequency and time. 
    
    \item The second approach is offline augmentation that generates extra data before training. 
    In the multi-stream framework, we conduct experiments with either simulated audio or real recordings from non-target streams. 
    In DIRHA \cite{ravanelli2016realistic}, several reverberated versions of clean speech are generated using pre-measured room impulse responses; In AMI \cite{carletta2005ami}, recordings from close-talk microphones in addition to microphone arrays are used for Stage-1 training. 
\end{itemize}

\subsubsection{Stage-2 Time Masking} 

Stage-2 augmentation aims to improve a multi-stream model's robustness against variations in inter-stream dynamics. For instance, the model needs to learn how to reliably handle the situation if one of the arrays suddenly fails in a meeting setting. 
Since the UFE features are the direct inputs for Stage-2, we consider augmentation on UFE features instead of log-Mel filter bank features. 

In this work, we introduce Stage-2 Time Masking, a simple but effective method to create differences across the streams. Inspired by temporal masking in SpecAugment, Stage-2 Time Masking masks the UFE features in time for individual streams. 
For each utterance during training, a pre-defined number of time masks are placed on the UFE features.
The mask will replace the value of the original UFE features with the filled mask value within the masking region. 
The applied location and duration of a mask are both randomly chosen from a uniform distribution. 
Note that Stage-2 Time Masking is applied only during training. 
The time mask is utterance-specific, in that it replaces the features with the mean value of the UFE features for that utterance.

The Stage-2 Time Masking is intended to mimic the situation of a partial loss of a speech segment for one of the streams. 
Compared with augmentation at the acoustic level, Stage-2 Time Masking is computationally easy to apply with no additional data.

\subsection{Adaptive CTC Fusion}

In the previous study \cite{li2020practical}, the CTC component of each stream was pre-trained in Stage-1 and kept frozen in Stage-2 for training. 
During inference in a multi-stream setting, equal decoding weights across all streams were assigned to the CTC components in Eq. \ref{eq:ctc}.
These pre-defined CTC weights could be problematic if one array is in an acoustic condition that is significantly worse than the others. 

In this work, we propose adaptive CTC fusion during decoding to mitigate the problem above using the knowledge from hierarchical attention mechanism. 
For every prediction, the hierarchical attention network produces an attention vector $[\beta_{l}^{(1)}, \beta_{l}^{(2)}, ... , \beta_{l}^{(N)}]$ across all streams, which steers the system to more informative streams. 
Since a label-synchronous beam search is employed during inference, each CTC component produces a prefix score, $\alpha_{\textrm{ctc}^{(i)}}(h)$, for a hypothesized sequence $h$.
Instead of taking average of stream-specific prefix scores for overall CTC contribution of hypothesis $h$, we calculate the weighted average contributions from individual CTCs.
The stream attention vector can be combined with CTC prefix scores $\alpha_\textrm{ctc}(h)$ for a hypothesized sequence $h$:
\begin{equation}
    \alpha_\textrm{ctc}(h)={\sum}_{i=1}^{N}\beta_{l}^{(i)}*\alpha_{\textrm{ctc}^{(i)}}(h),
\end{equation}
where adaptive stream weight $\beta_{l}^{(*)}$ is applied to each CTC network and $l$ is the index of the latest prediction of hypothesis $h$.

\section{Data}

Two datasets, DIRHA English WSJ \cite{ravanelli2016realistic} and AMI meeting corpus \cite{carletta2005ami}, were used for experiments and analysis. 
\begin{figure}[!htbp]
  \centering
  \centerline{\includegraphics[width=3.5cm]{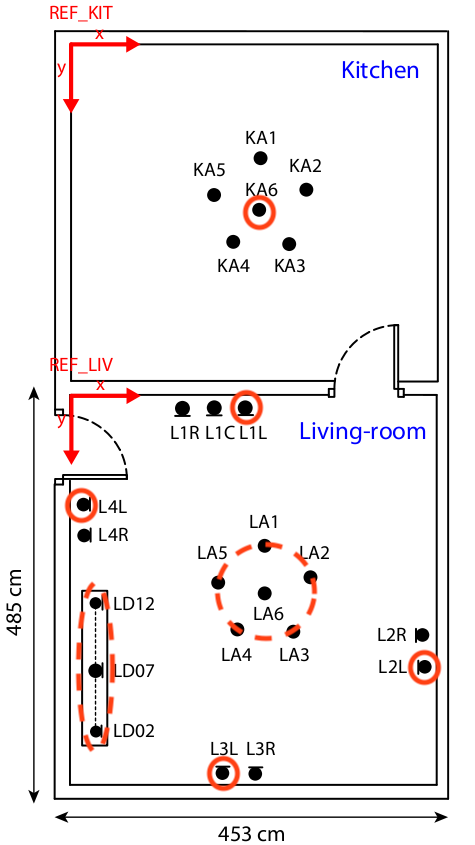}}
\caption{DIRHA English WSJ Microphone Configuration. Streams selected are in red circles. Beam Circular Array contains 6 microphones (LA1-LA6), Beam Linear Array includes 11 microphones (LD02-LD12).}
\label{fig:dirha}
\end{figure}

The DIRHA English WSJ corpus focuses on the challenge of speech interactions via distributed microphones in a domestic environment. 
There are in total 32 microphones placed in an apartment with a living room and a kitchen.
In our experiments, we chose two microphone arrays, Beam Linear Array (BLA) and Beam Circular Array (BCA), and five single microphones (depicted in Fig. \ref{fig:dirha}) for use in either training or evaluation. 
Training data was created by contaminating the original Wall Street Journal clean speech (WSJ0 and WSJ1, 81 hours in total) with room impulse responses for corresponding streams. 
The development set for cross validation was simulated with typical domestic background noise and reverberation. 
For evaluation, read WSJ utterances were newly recorded simultaneously by all 32 channels in a real setting.
In addition, we created a synthetic test stream, {\it NoMic}, to replicate the scenario of signal cut-off, where inputs are all zeros after mean and variance normalization.

The AMI meeting corpus was created in three instrumented rooms with meeting conversations. Each meeting room was configured with two microphone arrays and close-talk microphones for individual speakers, resulting in 100 hours of far-field signal-synchronized recordings. 
With segments of overlapping speakers removed, the training, development and evaluation set contain 81 hours, 9 hours and 9 hours of meeting recordings, respectively. 
Table \ref{tab:ami} summarizes the stream descriptions used in subsequent experiments. 
For stream {\it IHM}, the close-talk microphone with the most energy among all attendees was selected at each time frame. In contrast, stream {\it IHM0} always took speech from speaker-0, regardless of if speaker-0 was speaking.  
Similar to the DIRHA setup, {\it NoMic} was created to mimic constant microphone dropout. 

For each array in both datasets, multi-channel input was synthesized into a single-channel audio using the Delay-and-Sum beamforming technique with the BeamformIt Toolkit \cite{anguera2007acoustic}.

\begin{table}[!htbp]
  \begin{center}
  	\caption{AMI Meeting Corpus Stream Configuration.}
	\begin{tabular}{ll}
	  \toprule
	  \toprule
	  Stream & Description \\
	  \midrule
      MDM &	first microphone array \\
      SMDM	& second microphone array \\
      IHM & individual headset microphones\\
      IHM0 & individual headset microphones \\
      & (fixed speaker-0 for each meeting)\\
      NoMic	& constant stream dropout (all-zero inputs)\\
      \bottomrule
      \bottomrule
	\end{tabular}
    \label{tab:ami}
  \end{center}
\end{table}

\section{Experiment Setup}

All the experiments were conducted using the Pytorch backend on ESPnet \cite{watanabe2018espnet}. 
Table \ref{tab:config} describes the relevant setup information for the various experiments. 
Two model configurations were explored: 
{\it Config-1} included two BLSTM layers in the encoder and one LSTM layer in the decoder. 
A more complex model with {\it Config-2} had an additional two BLSTM layers and an extra LSTM layer as well. 
We used 50 distinct labels including 26 English letters and other special tokens, i.e., punctuation and sos/eos.
A look-ahead word-level RNN-LM \cite{hori2018end} was incorporated during inference.
It was trained separately using Stochastic Gradient Descent (SGD) for 20 epochs. 

\begin{table}[th]
  \begin{center}
  	\caption{Experimental Configuration.}
    \label{tab:config}
    \scalebox{0.8}{
	\begin{tabular}{ll}
	  \toprule
	  \toprule
	  {\bf Feature} & 80-dim log-mel filter bank + 3-dim pitch\\
      \midrule
	  {\bf Model}\\
      Encoder type & VGGBLSTM \cite{hori2017advances, cho2018multilingual} (subsampling factor: 4)\\
      Encoder layers & Config-1: 6(CNN)+2(BLSTM) \\
      & Config-2: 6(CNN)+4(BLSTM)\\
      Encoder units  & 320 cells (BLSTM layers)\\
      Encoder projection & 320 cells (BLSTM layers)\\
      Frame-level Attention & 320-cell Content-based\\
      Stream Attention & 320-cell Location-based\\
      Decoder type & LSTM\\
      Decoder layers & 1 (Config-1) or 2 (Config-2)\\
      Decoder units  & 320 cells\\
      \midrule 
	  {\bf Train and Decode} \\
	  Optimizer & AdaDelta\\
	  Batch size & 30 (Stage-1); 15 (Stage-2) \\
	  Training Epoch & 30 epochs (patience:3 epochs)\\
      CTC weight $\lambda$ & 0.2 (train); 0.3 (decode)\\
      Label Smoothing& Type: Unigram \cite{pereyra2017regularizing}, Weight: 0.05\\
      Beam size & 30\\
        \midrule 
      {\bf RNN-LM} \\
      Type& Look-ahead Word-level RNNLM \cite{hori2018end}\\
      Size & 1-Layer LSTM with 1,000 cells\\
      Vocabulary & 65,000 \\
      Train data & AMI:AMI; DIRHA:WSJ0-1+extra WSJ text\\
      LM weight $\gamma$ & AMI:0.5; DIRHA:1.0\\
        \midrule 
      {\bf SpecAugment \cite{Park2019SpecAugmentAS}} \\
      Time mask & \#masks: 2; $T$: 40 \\
      Frequency mask & \#masks: 2; $F$: 30\\
      \bottomrule 
      \bottomrule
	\end{tabular}}
  \end{center}
\end{table}

\section{Results and Discussions}

\subsection{Stage-1 Augmentation}

To investigate the effectiveness of Stage-1 augmentation, we evaluated online and offline augmentation techniques on DIRHA and AMI datasets.  
Table \ref{tab:stage1dirha} illustrates Stage-1 single-stream results using the proposed augmentation schemes. 
With the each model configuration, substantial Word Error Rate (WER) reductions were reported with SpecAugment, i.e., {\it  D1} v.s. {\it D3} and {\it D2} v.s. {\it D4}.  
Moreover, the more complex network {\it Config-2} did not necessarily improve over the smaller model {\it Config-1} until augmentation was utilized in training (i.e., {\it D1} outperformed {\it D2}, but {\it D4} outperformed all earlier models). 
We created additional reverberated copies of clean WSJ data using room impulse responses measured for four single microphones, i.e., {\it L1L}, {\it L2L}, {\it L3L} and {\it L4L}. 
{\it D11} achieved better WERs across six streams compared to {\it D5-D10}.
More importantly, {\it D11}, trained with all six streams, outperformed {\it D4} on the {\it BCA} and {\it BLA} evaluations, showing the value of the additional out-of-set data. From here, {\it D11} was selected as the Stage-1 model for the remaining DIRHA experiments. 

\begin{table}[!htbp]
  \begin{center}
   	\caption{Stage-1 Augmentation: DIRHA English WSJ. Model size (2, 1) and (4, 2) represent {\it Config-1} and {\it Config-2} in Table \ref{tab:config}. (\% WER)}
   	\scalebox{0.7}{
	\begin{tabular}{lccccccccc}
	  \toprule
	  \toprule
	   & Train&&Model&\multicolumn{6}{c}{Test Data}\\
	  ID&Data&SpecAug&Size&BCA&BLA&L1L&L2L&L3L&L4L\\
	   \midrule

      D1 & BCA+BLA & No & (2,1) & 33.9&30.7& --& --& --& --\\
      D2 & BCA+BLA & No & (4,2) & 34&32& --& --& --& -- \\
      D3 & BCA+BLA & Yes & (2,1) & 27.1& 24.4&--& --& --& -- \\
      D4 & BCA+BLA & Yes & (4,2) & 24.9&22.6& --& --& --& -- \\
      \midrule
      D5 & BCA & Yes & (4,2) & 27.1& -- & --& --& --& -- \\
      D6 & BLA & Yes & (4,2) & --& 27.7 & --& --& --& -- \\
      D7 & L1L & Yes & (4,2) & --& -- & 28.3& --& --& -- \\
      D8 & L2L & Yes & (4,2) & --& -- & --& 35.4& --& -- \\
      D9 & L3L & Yes & (4,2) & --& -- & --& --& 33& -- \\
      D10 & L4L & Yes & (4,2) & --& -- & --& --& --& 30.4 \\
      D11 & All Streams & Yes & (4,2) &\bf{19.8}&\bf{17.2}&\bf{22.6}&\bf{24.1}&\bf{22.6}&\bf{22.6}\\
      \bottomrule
      \bottomrule
	\end{tabular}
	}
    \label{tab:stage1dirha}
  \end{center}
\end{table}

Table \ref{tab:stage1ami} summarizes Stage-1 augmentation results of AMI in a similar way to Table \ref{tab:stage1dirha}.
It was clear looking at {\it A1-A4} that online augmentation (SpecAugment) consistently decreased error rates. 
Including additional close-talk stream {\it IHM}, {\it A8} showed lower WERs comparing to {\it A4}.
From here, {\it A8} was utilized for AMI Stage-2 training.

\begin{table}[!htbp]
  \begin{center}
   	\caption{Stage-1 Augmentation: AMI. (\% WER)}
   	\scalebox{0.7}{
	\begin{tabular}{lcccccc}
	  \toprule
	  \toprule
	   & Train&&Model&\multicolumn{3}{c}{Test Data}\\
	  ID&Data&SpecAug&Size&MDM&SMDM&IHM\\
	   \midrule

      A1 & MDM+SMDM & No & (2,1) & 56.9&61.7 & --\\
      A2 & MDM+SMDM & No & (4,2) & 53.1&58.3& --\\
      A3 & MDM+SMDM & Yes & (2,1)& 50.3& 54.9& --\\
      A4 & MDM+SMDM & Yes & (4,2) & 46.1 & 50.5 & --\\
      \midrule
      A5 & MDM & Yes & (4,2) & 50.5 & -- & --\\
      A6 & SMDM & Yes & (4,2) & --& 55.5 &--\\
      A7 & IHM & Yes & (4,2) & --& --& 30.4\\
      A8 & All Streams & Yes & (4,2) &\bf{42.8}&\bf{48.1}&\bf{27.6}\\
      \bottomrule
      \bottomrule
	\end{tabular}
	}
    \label{tab:stage1ami}
  \end{center}
\end{table}

\subsection{Adaptive CTC Fusion}

\subsubsection{Issues with Pre-defined CTC weights}

In previous study \cite{li2020practical}, each CTC network in the multi-stream setting contributed equally during inference. 
These pre-defined CTC weights could cause performance degradation if one of streams is corrupted.
We designed simple experiments in DIRHA to illustrate this issue. 
After Stage-1, we formulated a two-stream model using target streams, {\it BLA} and {\it NoMic} for training and testing. 
Since {\it BLA} was known to be the only informative source, stage-1 performance of $17.2\%$ for {\it BLA} was viewed as the best possible result. 
In Table \ref{tab:sfctc}, the {\it Oracle} Stage-2 decoding setup with CTC weights $[1.0; 0.0]$ achieved WER of $17.3\%$, essentially equivalent to the single-stream performance.
However, WER increased to $20.5\%$ when equal weights were applied. 
The proposed adaptive CTC fusion made the model more robust with the help of stream attention, reaching Stage-1 performance of $17.2\%$ without any pre-existing knowledge of the relative value of the streams. 

\begin{table}[!htbp]
  \begin{center}
  	\caption{Issues with Pre-defined CTC Weights. (\% WER)}
  	\scalebox{1}{
	\begin{tabular}{lc}
	  \toprule
	  \toprule
      Model & Test \\
	   \midrule
	   \it{Stage-1: BLA only}\\
	   D11 in Table \ref{tab:stage1dirha} & 17.2 \\ 
	   \midrule
	   \it{Stage-2: BLA-NoMic}\\
	   Pre-defined CTC Weights [1.0; 0.0] & 17.3 \\ 
	   Pre-defined CTC Weights [0.5; 0.5] & 20.5 \\ 
	   Adaptive CTC Fusion & \bf{17.2} \\ 
      \bottomrule
      \bottomrule
	\end{tabular}
	}
    \label{tab:sfctc}
  \end{center}
\end{table}

\subsubsection{Adaptive CTC Fusion: Matched Condition}

To show the influence of adaptive CTC fusion in matched conditions, we conducted experiments with different two-stream acoustic conditions.
In each experiment, training and evaluation data were drawn from the same arrays. 
Results are displayed in Table \ref{tab:match}.
In order to pick diverse conditions in DIRHA, three two-stream configurations were chosen, {\it BLA-L2L}, {\it BLA-BCA} and {\it L3L-L4L}. 
According to the Stage-1 performance, {\it BLA} was most informative single stream. {\it BCA}/{\it L2L} were the most similar/different streams to {\it BLA} in terms of WER. 
{\it L3L} and {\it L4L} resulted the same WER of $22.6\%$.
For AMI, all three conbinations of the three streams were selected. 
WER improvements were observed across all six cases in the two datasets. 
From the analysis of the case {\it BLA-L2L}, we observed increasing percentage of improved utterances ($77.0\%\rightarrow 78.5\%$). Note that the case of improved utterances describes the situation where WER from a multi-stream model is the same as or lower than the best single stream WERs. 

\begin{table}[!htbp]
  \begin{center}
  	\caption{Adaptive CTC Fusion in Matched Conditions. (\% WER)}
  	\scalebox{0.75}{
	\begin{tabular}{lccc}
    	  \toprule
    	  \toprule
           Decoding Strategy& \multicolumn{3}{c}{Train/Test Data}\\
    	   \midrule
           {\bf DIRHA}&  {\it BLA-L2L}& {\it BLA-BCA}& {\it L3L-L4L}\\
    	   \midrule
    	   Pre-defined CTC [0.5; 0.5] & 17.2&16.5&20.4\\
    	   Adaptive CTC Fusion & \bf{16.9}&\bf{16.1}&\bf{20.1} \\
    	   \midrule
           {\bf AMI}&  {\it MDM-SMDM} & {\it MDM-IHM} & {\it SMDM-IHM} \\
    	   \midrule
    	   Pre-defined CTC [0.5; 0.5] &42&29.3&29.8\\
    	   Adaptive CTC Fusion &\bf{41.6}&\bf{28.2}&\bf{28.3}\\
          \bottomrule
          \bottomrule
	\end{tabular}
	}
    \label{tab:match}
  \end{center}
\end{table}

\subsubsection{Adaptive CTC Fusion: Mismatched Condition}

For the following experiments, we designated {\it BLA-L2L} and {\it MDM-SMDM} as the training stream configurations for DIRHA and AMI, respectively. 
In DIRHA, three mismatched test conditions were chosen:  
{\it BLA-NoMic} and {\it BLA-KA6} were the unseen scenarios where one stream ({\it BLA}) is known to greatly outperform the other.
Note that {\it KA6} (Stage-1 WER: $61\%$) was a microphone in the kitchen while speakers read in the living room;
{\it L3L-L4L} were the microphones with the same Stage-1 performances. 
We specified two mismtached condtions for AMI: {\it MDM-NoMic} and {\it MDM-IHM0}. Recall {\it IHM0} (Stage-1 WER: $73.7\%$) is the close-talk microphone attached to speaker-0. 
In DIRHA, results in Table \ref{tab:mismatch} reported moderate improvement except {\it BLA-NoMic}, which sees a modest decline.
Stream {\it NoMic} is an extreme case and may be too aggressive as a unseen test stream.  
For AMI, relative WER reductions of $6.5\%$ and  $4.8\%$ were shown for the mismatched conditions. 

\begin{table}[!htbp]
  \begin{center}
  	\caption{Adaptive CTC Fusion in Mismatch Conditions. (\% WER)}
  	\scalebox{0.75}{
	\begin{tabular}{lccc}
    	  \toprule
    	  \toprule
           Decoding Strategy& \multicolumn{3}{c}{Test Data}\\
    	   \midrule
           {\bf DIRHA (BLA-L2L)}&  {\it BLA-NoMic}& {\it BLA-KA6}& {\it L3L-L4L}\\
          \midrule
    	   Pre-defined CTC [0.5; 0.5] & \bf{26.9}&21&20.3\\
    	   Adaptive CTC Fusion & 27.1&\bf{20.7}&\bf{20} \\
    	   \midrule
           {\bf AMI (MDM-SMDM)}&  {\it MDM-NoMic} & {\it MDM-IHM0} & -- \\
    	   \midrule

    	   Pre-defined CTC [0.5; 0.5] &46.1&44&--\\
    	   Adaptive CTC Fusion &\bf{43.1}&\bf{41.9}&--\\
          \bottomrule
          \bottomrule
	\end{tabular}
	}
    \label{tab:mismatch}
  \end{center}
\end{table}

\subsection{Stage-2 Time Masking}

To demonstrate another potential weakness of the previous MEM-Array system, we designed experiments in DIRHA to demonstrate potential performance degradation because of a mismatched test condition, as depicted in Table \ref{tab:toy}. 
{\it BLA-L2L} and {\it BLA-NoMic} were used to train and test two Stage-2 models.
While the matched conditions on the diagonal of Table \ref{tab:toy} exhibited reasonable results, the model trained with {\it BLA-L2L} is unable to handle the unseen condition {\it BLA-NoMic}, degrading by nearly $10\%$ absolute WER decrease comparing to Stage-1 {\it BLA} performance, $17.2\%$. 

\begin{table}[!htbp]
  \begin{center}
  	\caption{Evaluation in Matched and Mismatched Conditions. (\% WER)}
  	\scalebox{1}{
	\begin{tabular}{lcc}
    	  \toprule
    	  \toprule
    	    & \multicolumn{2}{c}{Test Data}\\
           Model & {\it BLA-L2L} & {\it BLA-NoMic}\\
    	   \midrule
    	   Stage-2 {\it BLA-L2L}  & 16.9 & \bf{27.1} \\
    	   Stage-2 {\it BLA-NoMic} & 17.6 & 17.2 \\
          \bottomrule
          \bottomrule
	\end{tabular}
	}
    \label{tab:toy}
  \end{center}
\end{table}

Stage-2 Time Masking was proposed to target the scenario mentioned above with results in Table \ref{tab:tm}.
Two multi-stream models were trained using {\it BLA-L2L} (DIRHA) and {\it MDM-SMDM} (AMI), respectively. 
During training, a time mask was created with the length uniformly sampled from $[0, 10]$ (in frames). 
Note that 10 frames accounted for 0.4 second due to subsampling. 
The mask was applied in a randomly selected position on the UFE features. 

We experimented with different numbers of time masks.
For DIRHA experiments, a model trained with 3 time masks per stream gave the optimal results. 
In particular, substantial absolute WER improvement of $9.3\%$ were seen when evaluating {\it BLA-NoMic}, presumably because it is essentially the situation that stage-2 augmentation is simulating. 
WER on {\it BLA-KA6} also decreased while keeping other conditions unchanged or slightly improved. 
In AMI experiments, Stage-2 augmentation kept all test conditions under similar performances.
It is likely that the AMI model could already handle these unseen conditions properly. 
For instance, without Stage-2 time masking, {\it MDM-NoMic} achived a WER of $43.1\%$  which is close to the Stage-1 {\it MDM} performance of $42.8\%$.
The number of masks is set to be 3 for DIRHA and 1 for AMI based on matched condition performances. 

For comparison, input dropout on the UFE features was implemented. Stage-2 Time Masking constantly obtained lower WERs in all conditions, which supported the idea of creating stream dynamics instead of unit dropout over the inputs.

\begin{table}[!htbp]
  \begin{center}
  	\caption{Stage-2 Time Masking. (\% WER)}
  	\scalebox{0.7}{
	\begin{tabular}{lcccc}
    	  \toprule
    	  \toprule
           Model & \multicolumn{4}{c}{Test Data}\\
    	   \midrule
           {\bf DIRHA}&  {\it BLA-L2L}& {\it BLA-NoMic}&{\it BLA-KA6}& {\it L3L-L4L}\\
    	   \midrule
            Stage2 BLA-L2L &  16.9 & 27.1 & 20.7 &20 \\
             - Input Dropout 0.2 & 17.7 &38.1 &22.1& 20.6 \\
             - Input Dropout 0.5 &19.2 &21 &23.6 &22.6 \\
             - Time Masking (\#mask=1) & 17 &	18&	19.3&	20.1 \\
             - Time Masking (\#mask=2) &16.9&	18.2&	19.4&	20\\
             - Time Masking (\#mask=3) &\bf{16.6}&	\bf{17.8}&\bf{	19.2}&	\bf{20}\\
    	   \midrule

           {\bf AMI}&  {\it MDM-SMDM}& {\it MDM-NoMic}&{\it MDM-IHM0}& --\\
    	   \midrule
            Stage2 MDM-SMDM & 41.6&	43.1&	41.9&--\\
             - Input Dropout 0.2 & 42.3&	44.5&	42.6&--\\
             - Input Dropout 0.5 & 45.2&	49.5&	46.3&--\\
             - Time Masking (\#mask=1) &  \bf{41.6}&	\bf{43.1}&	\bf{41.6}&--\\
             - Time Masking (\#mask=2) &41.6&	43&	41.9&--\\
             - Time Masking (\#mask=3) & 41.7&	43&	41.8&--\\
          \bottomrule
          \bottomrule
	\end{tabular}
	}
    \label{tab:tm}
  \end{center}
\end{table}


\subsection{Discussion on Amount of Parallel Data}

Generally, parallel data are more expensive to collect. 
In this section, we examined how much parallel data could be sufficient for Stage-2 model training with a reasonable performance. 
We used {\it BLA-L2L} in DIRHA for this demonstration.
As described in Table \ref{tab:datasize}, 1 hour data per stream could maintain fair WERs with only an average of $5.3\%$ performance degradation, indicating a relatively low burden for data resources. 

\begin{table}[!htbp]
  \begin{center}
  	\caption{Discussion on Amount of Parallel Data. (\% WER)}
  	\scalebox{0.9}{
	\begin{tabular}{lcccc}
    	  \toprule
    	  \toprule
           Training Data  & \multicolumn{4}{c}{Test Data}\\
           (Hours)&  {\it BLA-L2L}& {\it BLA-NoMic}&{\it BLA-KA6}& {\it L3L-L4L}\\
    	   \midrule
            0.1 & 17.1&	24&	21.3&	20.4\\
             1 & 17&	20.4&	20&	20.1\\
             10 &16.7&	18.1&	18.9&	20.2\\
             81 (All) & 16.6&	17.8&	19.2&	20.1\\
          \bottomrule
          \bottomrule
	\end{tabular}
	}
    \label{tab:datasize}
  \end{center}
\end{table}

\subsection{Overall}

Table \ref{tab:overall} summarizes the contributions of each proposed step, in this case using 
{\it BLA-L2L} and {\it MDM-SMDM} as the training stream configurations for DIRHA and AMI, respectively, while test data includes matched and mismatched conditions.
Stage-1 augmentation together with a more complex model consistently reduced the WERs. 
Adaptive CTC fusion and Stage-2 time masking provided notable improvements in various scenarios. 
Overall, compared to the previous training strategy \cite{li2020practical}, we observed average relative WER reductions of $45.2\%$ (DIRHA) and $30.7\%$ (AMI). 
In particular, substantial relative WER improvement of $29.7-59.3\%$ was reported across several mismatched stream conditions. 
For fair comparison, we also evaluated the model where the HAN component was replaced by fixed stream fusion weights $[0.5;0.5]$ for fusion of context vectors.  
In these cases, the components, including CTC, frame-level attention and decoder, were optimized during Stage-2. 
Our proposed model greatly outperformed the model with no stream attention.

\begin{table}[!htbp]
  \begin{center}
  	\caption{Overall Results. (\% WER)}
  	\scalebox{0.7}{
	\begin{tabular}{lcccc}
    	  \toprule
    	  \toprule
           Model & \multicolumn{4}{c}{Test Data}\\
    	   \midrule
           {\bf DIRHA (BLA-L2L)}&  {\it BLA-L2L}& {\it BLA-NoMic}&{\it BLA-KA6}& {\it L3L-L4L}\\
    	   \midrule
            Two-Stage Training & 27.4&	43.8&	37.9&	29.7\\
             + Large Model & 28&	57.4&	37.8&	29.7\\
             + Stage-1 Augment. &17.2&	26.9&	21&	20.3\\
             + Adaptive CTC Fusion & 16.9&	27.1&	20.7&	20\\
             + Stage-2 Time Masking &\bf{16.6(39.4\%)}&	\bf{17.8(59.3\%)}&\bf{19.2(49.3\%)}&	\bf{20(32.7\%)}\\
    	   No Stream Attention & 36.2&	66.5&	49.4&	37.2\\
    	   \midrule
           {\bf AMI (MDM-SMDM)}&  {\it MDM-SMDM}& {\it MDM-NoMic}&{\it MDM-IHM0}& --\\
    	   \midrule
            Two-Stage Training& 55.5&	69&	59.2&--\\
             + Large Model & 52&	62&	55.1&--\\
             + Stage-1 Augment. &42&	46.1&	44&--\\
             + Adaptive CTC Fusion & 41.6&	43.1&	41.9&--\\
             + Stage-2 Time Masking &\bf{41.6(25.0\%)}&	\bf{43.1(37.5\%)}&	\bf{41.6(29.7\%)}&--\\
    	   No Stream Attention & 56&	69.7&	65.8&--\\
          \bottomrule
          \bottomrule
	\end{tabular}
	}
    \label{tab:overall}
  \end{center}
\end{table}

\newcommand{\ww}{3cm}
\newcommand{\pos}{0.32}
\begin{figure}[h!]

\begin{minipage}[b]{\pos\linewidth}
  \centering
  \centerline{\includegraphics[width=\ww]{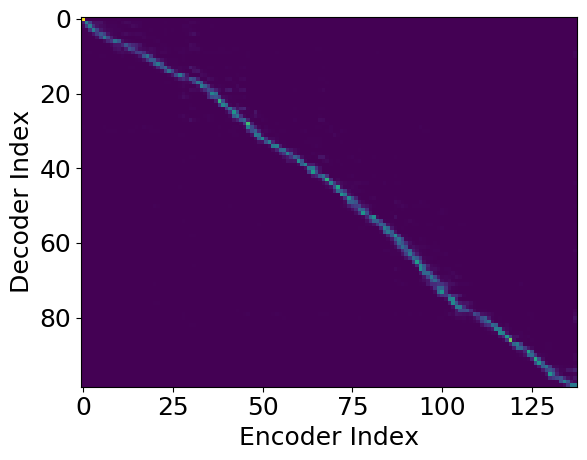}}
 \vspace{-0.1cm}
  \centerline{\scriptsize{(a) MDM}}\medskip
\end{minipage}
\hfill
\begin{minipage}[b]{\pos\linewidth}
  \centering
  \centerline{\includegraphics[width=\ww]{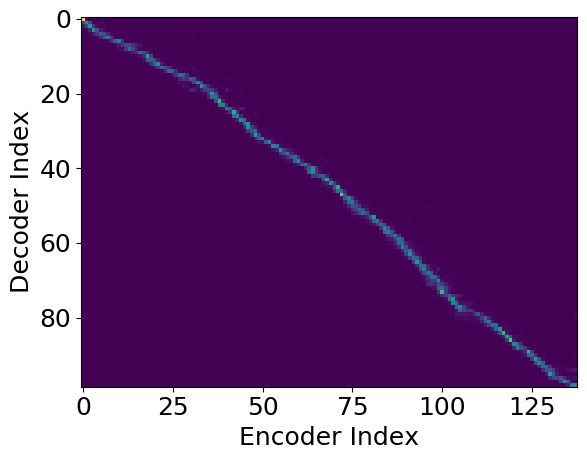}}
 \vspace{-0.1cm}
  \centerline{\scriptsize{(b) IHM0}}\medskip
\end{minipage}
\hfill
\begin{minipage}[b]{\pos\linewidth}
  \centering
  \centerline{\includegraphics[width=\ww]{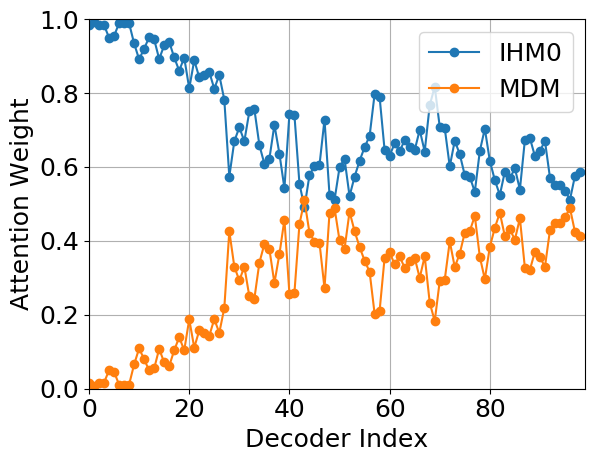}}
 \vspace{-0.1cm}
  \centerline{\scriptsize{(c) MDM-IHM0}}\medskip
\end{minipage}

\begin{minipage}[b]{\pos\linewidth}
  \centering
  \centerline{\includegraphics[width=\ww]{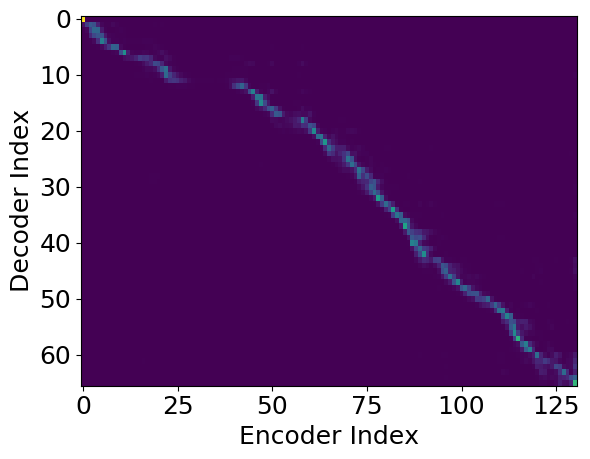}}
 \vspace{-0.1cm}
  \centerline{\scriptsize{(d) MDM}}\medskip
\end{minipage}
\hfill
\begin{minipage}[b]{\pos\linewidth}
  \centering
  \centerline{\includegraphics[width=\ww]{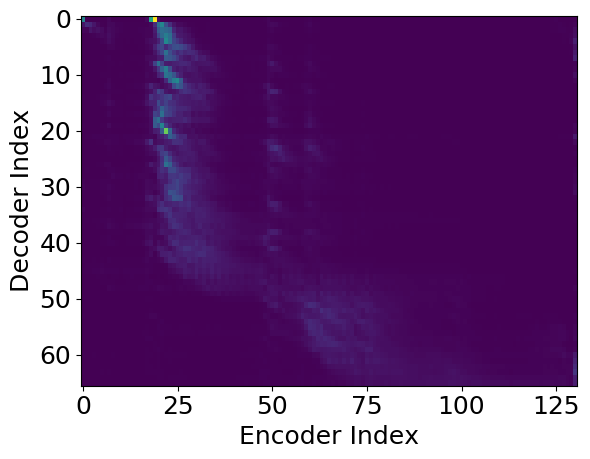}}
 \vspace{-0.1cm}
  \centerline{\scriptsize{(e) IHM0}}\medskip
\end{minipage}
\hfill
\begin{minipage}[b]{\pos\linewidth}
  \centering
  \centerline{\includegraphics[width=\ww]{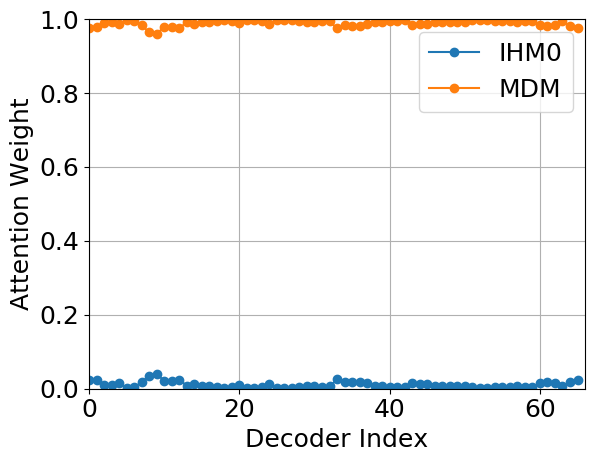}}
 \vspace{-0.1cm}
  \centerline{\scriptsize{(f) MDM-IHM0}}\medskip
\end{minipage}

\caption{Sentence Analysis of Attention Mechanism during Inference. Example 1 (speaker-0 speaking) includes (a),(b),(c); Example 2 (speaker-0 not speaking) includes (d),(e),(f). (a) and (d) are frame-wise attention alignments of {\it MDM}; (b) and (e) are frame-wise attention alignments of {\it IHM0}; (c) and (f) are stream attention weights of {\it MDM-IHM0}.}
\label{fig:res}
\end{figure}

To visualize the effect of the stream attention, Fig. \ref{fig:res} shows attention plots of two examples from evaluation set {\it MDM-IHM0} in AMI. 
In the first example, (a)-(c), speaker-0 was speaking, and as a result both {\it MDM} and {\it IHM0} were informative sources, and the stream attention in (c) gave weights to both inputs, though shifted slightly towards {\it IHM0} since this close-talk stream had better speech quality. 
In the second example, (d)-(f), speaker-0 was not speaking and so another speaker's audio was recorded by {\it MDM} while {\it IHM0} could barely capture any speech. 
In this case, the stream fusion mechanism correctly attended to {\it MDM} with nearly $100\%$ confidence.

\section{Conclusion}

In this work, we presented a two-stage augmentation scheme and adaptive CTC fusion for the purpose of improving robustness of the multi-stream end-to-end model against diverse testing conditions. 
Inherited from the two-stage training strategy, the two-stage augmentation consistently improved performance across matched and mismatched conditions; adaptive CTC fusion enhances the robustness by applying stream attention weights dynamically. 
For future research, stream-specific knowledge could be used for a more customized stage-2 training, and more sophisticated attention mechanisms could be explored for stream fusion.



\bibliographystyle{IEEEbib}
\bibliography{strings,refs}

\end{document}